\documentclass[%
 aip,
 amsmath,amssymb,
preprint,reprint, 
]{revtex4-1}

\usepackage{graphicx}
\usepackage{dcolumn}
\usepackage{bm}
\usepackage{amsmath,stackengine,scalerel,calrsfs}
\usepackage[hidelinks, linkcolor=blue, citecolor=blue, colorlinks=true]{hyperref}

\usepackage[utf8]{inputenc}
\usepackage[T1]{fontenc}
\usepackage{mathptmx}
\usepackage{etoolbox}
\usepackage{enumitem}
\DeclareMathAlphabet{\pazocal}{OMS}{zplm}{m}{n}
\stackMath
\newcommand\reallywidehat[1]{%
\savestack{\tmpbox}{\stretchto{%
  \scaleto{%
    \scalerel*[\widthof{\ensuremath{#1}}]{\kern-.6pt\bigwedge\kern-.6pt}%
    {\rule[-\textheight/2]{1ex}{\textheight}}
  }{\textheight}%
}{0.5ex}}%
\stackon[1pt]{#1}{\tmpbox}%
}
\makeatletter
\def\@email#1#2{%
 \endgroup
 \patchcmd{\titleblock@produce}
  {\frontmatter@RRAPformat}
  {\frontmatter@RRAPformat{\produce@RRAP{*#1\href{mailto:#2}{#2}}}\frontmatter@RRAPformat}
  {}{}
}%
\makeatother

\newcommand{\bB}{\textit{\textbf{B}}}

\newcommand{\bb}{\textit{\textbf{b}}}

\newcommand{\bJ}{\textit{\textbf{J}}}
\newcommand{\bj}{\textit{\textbf{j}}}
\newcommand{\bk}{\textit{\textbf{k}}}
\newcommand{\bp}{\textit{\textbf{p}}}
\newcommand{\bq}{\textit{\textbf{q}}}
\newcommand{\bu}{\textit{\textbf{u}}}
\newcommand{\bx}{\textit{\textbf{x}}}
\newcommand{\by}{\textit{\textbf{y}}}
\newcommand{\bz}{\textit{\textbf{z}}}

\newcommand{\bw}{{\boldsymbol{\omega}}}

\newcommand{\bnabla}{{\boldsymbol{\nabla}}}

\begin{document}
\preprint{AIP/123-QED}
\title{Fundamental units of triadic interactions in Hall magnetohydrodynamic turbulence: how far can we go?}
\author{Supratik Banerjee}%
\author{Arijit Halder}
\email{sbanerjee@iitk.ac.in}
\affiliation{Department of Physics, Indian Institute of Technology Kanpur}

\date{\today}
\begin{abstract}
A systematic study has been carried out to obtain the fundamental units of triad interaction in Hall magnetohydrodynamic turbulence. Instead of finding the elementary building blocks of non-unique mode-to-mode transfer rates, we have investigated the fundamental units for uniquely defined combined transfers and convincingly showed that the mode-to-mode transfers can act as a practical base element for the same. In addition to the conventional field-specific mode-to-mode transfers, here we have introduced the idea of mode-specific transfers which is found to be important for the turbulent cascade and the turbulent relaxed states. Whereas the Hall transfer is found to associate mode-to-mode transfers for mode-specific interactions (with a three-member basis), it presents a mixture of typical mode-to-mode (also with a three-member basis) and non mode-to-mode (with a five-member basis) transfers for the field-specific interactions. The non mode-to-mode transfers are shown to satisfy the triad conservation differently from the mode-to-mode transfers. However, they also possess an inherent non-uniqueness and hence cannot be determined unambiguously unlike the combined transfer rates.
\end{abstract}

\maketitle

\section{Introduction}
Turbulence is a complex flow regime dominated by nonlinearities \cite{Frisch_1995, Tennekes_2018, Pope_2001}. For neutral fluids, this nonlinearity is represented by the velocity advection term $\left(\bu\times\bw\right)$ where $\bu$ is the fluid velocity and $\bw=\bnabla\times\bu$\citep{Lamb_1877,Banerjee_2016c, Andres_2019b}. In case of a magnetohydrodynamic (MHD) fluid, the nonlinearity in momentum equation is represented by both $\left(\bu\times\bw\right)$ and the Lorentz force term $\left(\bJ\times\bB\right)$, where $\bB$ is the magnetic field and $\bJ= \mu_0^{-1} \left(\bnabla \times \bB \right) $ is the current density with $\mu_0$ being the free-space permeability \citep{Biskamp_2003, Galtier_2016}. In case of ordinary MHD turbulence, the induction equation takes the form 
\begin{equation}
\frac{\partial \bB}{\partial t} = \bnabla \times \left(\bu \times \bB\right)+\eta\nabla^2 \bB,
\end{equation} 
where the first term on the right hand side represents the nonlinear contribution and the second term represents the diffusion, $\eta$ being the magnetic diffusivity. However, the ordinary MHD model breaks down when one tries to probe length scales comparable to the ion inertial length scale $d_i$. The simplest mono-fluid model to study such a plasma is given by Hall magnetohydrodynamics (HHMD). The Hall effect is modelled in the induction equation by adding a nonlinear term $- \bnabla \times \left(\bJ \times \bB\right)/ne$ where $n$ and $e$ are the number density of the electrons or ions and the electronic charge respectively. In terms of $\bb$ ($= \bB/\sqrt{\mu_0 \rho}$ with $\rho$ being the mass density of MHD fluid), and $\bj= \bnabla \times \bb$, the induction equation simply gives the evolution of $\bb$ and corresponding Hall term can be written as $-d_i\bnabla \times\left(\bj \times \bb\right)$. Similar to MHD, the total energy $E=\int(u^2+b^2)/2\ d\tau$ is a quadratic inviscid invariant in HMHD thereby leading to a nonlinear cascade of energy in the inertial range\footnote{A range of intermediate length scales unaffected by both large-scale forcing and small-scale dissipation}. Evidently, the nonlinear terms in both momentum and induction equations are responsible for the effective nonlinear transfer of the corresponding energy cascade\citep{Banerjee_2013,Banerjee_2018,Banerjee_2020, Banerjee_2016c,Verma_2021}. In particular, such transfers can be shown to occur due to the interaction of wave vector triads  $({\bk}, \bp, \bq)$, such that $\bk + \bp + \bq= {\bf 0}$ and the wave vectors are of comparable size \citep{Waleffe_1992,Verma_2004, Alexakis_2005,Mininni_2005b, Domaradzki_1990}. Interestingly, for each triad, the conservation of energy also implies
\begin{equation}
  S(\bk|\bp,\bq)+S(\bp|\bq,\bk)+S(\bq|\bk,\bp)=0  \label{triad}
\end{equation}
where $S(\bk|\bp,\bq)$ is interpreted as the combined transfer rate of energy to the $\bk$-th mode jointly from $\bp$ and $\bq$-th mode. Such triads practically constitute the fundamental units of energy conservation in spectral space. Such triadic conservations are found to exist both in incompressible hydrodynamic (HD) and MHD turbulence with and without rotation \citep{Kraichnan_1958,Sagaut_2008,Lesieur_1987}. The combined transfer rate $S(\bk|\bp,\bq)$ can further be decomposed as a sum of two mode-to-mode (M2M hereafter) transfers $S(\bk|\bp|\bq)$ and $S(\bk|\bq|\bp)$, where $S(\bk|\bp|\bq)$ represents the energy flux rate to the $\bk$-th mode from $\bp$-th mode with $\bq$-th mode as the mediator and $S(\bk|\bq|\bp)$ represents the energy flux rate from the $\bq$-th mode with $\bp$-th mode as the mediator \citep{Kraichnan_1959, Verma_2004}. $S(\bk|\bp|\bq)$, by definition, is anti-symmetric under the giver-receiver permutation in a triad \textit{i.e.,} $S(\bk|\bp|\bq)=-S(\bp|\bk|\bq)$ and can only be determined up to an arbitrary circulation function $\Delta(\bk|\bp|\bq)$ such that $\Delta(\bk|\bp|\bq)+\Delta(\bk|\bq|\bp)=0$ \citep{DAR_2001, Plunian_2019}. Despite being nonunique, M2M transfer rates are believed to be the building blocks of triadic interactions and a knowledge of M2M transfer rates computationally facilitates the study of turbulent transfers. For example, in dynamo action, the M2M transfer rates are used to calculate the scale-specific fluxes responsible for large-scale magnetic field growth both in MHD and HMHD turbulence\citep{Kumar_2013,Kumar_2017,Halder_2023}. 

In a recent work, \citet{Plunian_2019} investigated the uniqueness of M2M transfers for both incompressible HD and MHD turbulence. For each M2M transfer and the corresponding circulation transfer, they defined a basis consisting of scalar triple products of Fourier amplitudes of field variables (velocity, vorticity, magnetic field, current etc.). Using that specific form of basis they showed that, unlike pure hydrodynamics, the M2M transfer rates associated with the magnetic field can be uniquely determined if one decomposes the nonlinear term in the induction equation as $\bnabla\times(\bu\times\bb)=-(\bu\cdot\bnabla)\bb+(\bb\cdot\bnabla)\bu$ and respectively associate them with the advection and the stretching from physical consideration. 

In this paper, we argue that such uniqueness of M2M transfer rates in MHD is not mathematically justified and appears only due to the improper choice of basis. In fact, by construction, all the M2M transfers of MHD are non-unique up to a circulation function and hence the unambiguous determination of the number of the base functions is impossible. It is rather meaningful to define a set of base functions for the combined transfer rates which are uniquely defined in a triad. For HD and MHD, we obtain two types of bases depending on whether the combined transfer rates are expressed in a `mode-specific' or in a `field-specific' manner. Finally, we implement our methodology to find the base functions of different nonlinear transfers due to the Hall term in HMHD turbulence and provide a plausible phenomenological picture corresponding to it. Our work, for the first time, provides the complete set of base functions which constitute the family of building blocks for the nonlinear transfers in HMHD.  

\section{Governing equations and nonlinear transfers}
To systematically analyse the nonlinear transfers, we start with the equations of incompressible HMHD: 
\begin{align}
\partial_t \bu -  \nu\nabla^2 {\bu} &= -\left(\bw\times\bu\right) +  \left(\bj\times\bb\right)  -\bnabla P + {\textbf{\textit f}} , \label{eq:u1}\\
\partial_t \bb - \eta\nabla^2 {\bb} &= \bnabla\times\left(\bu\times\bb\right) - d_i\bnabla \times\left(\bj \times \bb\right),\label{eq:b1}\\
\bnabla \cdot \bu &= 0, \;\;   \bnabla \cdot \bb =  0, \label{eq:div}
\end{align}
where $P\ (=p+u^2/2)$ is the total pressure, {\textbf{\textit f}} is an external force, $\nu$ and $\eta$ denote the kinematic viscosity and the magnetic diffusivity respectively. As mentioned earlier, the last term on the right hand side of Eq.~\eqref{eq:b1} corresponds to the Hall term \citep{Galtier_2016, Bittencourt_2013}. Note that, for incompressible turbulence, all the above nonlinear terms can be written in various ways-
\begin{align}
    \bw\times\bu&=(\bu\cdot\bnabla)\bu-\bnabla\left(\frac{u^2}{2}\right)=\bnabla\cdot\left(\bu\otimes\bu\right),\label{NL1}\\
    \bj\times\bb&=(\bb\cdot\bnabla)\bb-\bnabla\left(\frac{b^2}{2}\right)=\bnabla\cdot\left(\bb\otimes\bb\right),\label{NL2}\\
    \bnabla\times(\bu\times\bb)&=(\bb\cdot\bnabla)\bu-(\bu\cdot\bnabla)\bb=\bnabla\cdot\left(\bb\otimes\bu-\bu\otimes\bb\right),\label{NL3}\\
    \bnabla\times(\bj\times\bb)&=(\bb\cdot\bnabla)\bj-(\bj\cdot\bnabla)\bb=\bnabla\cdot\left(\bb\otimes\bj-\bj\otimes\bb\right).\label{NL4}
\end{align}
All the gradient terms can be absorbed in the pressure term and do not finally contribute to the total flux of energy $E=\int(u^2+b^2)/2\ d\tau$ \citep{Banerjee_2016a, Banerjee_2016b, Banerjee_2023}. As we shall see, different expressions can be associated with different nature of modal transfers of energy. 
For the cascade of total energy, the inertial range modal energy density (for $\bk$-th mode) evolves as 
\begin{align}
&\partial_t  \left(E^u_\bk+E^b_\bk\right)\label{Energy_Fourier_timeu2} \\ 
&= \frac{1}{2}  \int\limits_{\bk + \bp + \bq= {\bf 0}}\!\!\!
\left[ S^{K}\left(\bk|\bp,\bq\right)+S^{M}\left(\bk|\bp,\bq\right)+ d_i S^{\pazocal{H}}\left(\bk|\bp,\bq\right) \right] d {\bp},\nonumber
\end{align}
where $E^u_\bk = \left(\bu_\bk\cdot\bu_\bk^*\right)/2$, $E^b_\bk = \left(\bb_\bk\cdot\bb_\bk^*\right)/2$,
\begin{align}
S^{K}\left(\bk|\bp,\bq\right)&=\Re\{\bu_\bk\cdot\left(\bu_\bp\times\bw_\bq\right)+\bu_\bk\cdot\left(\bu_\bq\times\bw_\bp\right)\},\label{SE1}\\
S^{M}\left(\bk|\bp,\bq\right)&=\Re\{\bu_\bk\cdot\left(\bj_\bp\times\bb_\bq\right)+\bu_\bk\cdot\left(\bj_\bq\times\bb_\bp\right)\nonumber\\ 
&+\bj_\bk\cdot\left(\bu_\bp\times\bb_\bq\right)+\bj_\bk\cdot\left(\bu_\bq\times\bb_\bp\right)\},\label{SE2}\\
S^{\pazocal{H}}\left(\bk|\bp,\bq\right)&=\Re\{\bj_\bk\cdot(\bb_\bq\times\bj_\bp)+\bj_\bk\cdot(\bb_\bp\times\bj_\bq)\},\label{SE3}
\end{align}
$\Re$ denotes the real part of a complex number and for any quantity $\psi$, $\psi_{\bk} \equiv \reallywidehat{\psi} \left(\bk\right)$. 

\section{mode-specific M2M transfer rates}
By definition, $S^{K}\left(\bk|\bp,\bq\right)$, $S^{M}\left(\bk|\bp,\bq\right)$ and $S^{\pazocal{H}}\left(\bk|\bp,\bq\right)$ are symmetric in $(\bp, \bq)$ and denote the combined transfer rates of energy to the $\bk$-th mode involving the velocity modes only, involving both velocity and magnetic modes and due to the Hall term respectively. After a careful consideration of Eqs.~\eqref{SE1}-\eqref{SE3} and the basic properties of M2M transfers (as mentioned before), one can effectively write the corresponding M2M transfers (choosing the circulation functions to be identically zero) as 
\begin{align}
    S^{K}(\bk|\bp|\bq)&=\Re\{\bu_\bk\cdot\left(\bu_\bp\times\bw_\bq\right)\},\label{m2mK}\\
    S^{M}(\bk|\bp|\bq)&=\Re\{\bu_\bk\cdot\left(\bj_\bp\times\bb_\bq\right)+\bj_\bk\cdot\left(\bu_\bp\times\bb_\bq\right)\},\label{m2mM}\\
    S^{\pazocal{H}}(\bk|\bp|\bq)&=\Re\{\bj_\bk\cdot(\bb_\bq\times\bj_\bp)\},\label{m2mH}
\end{align}
and similarly obtains the other M2M transfers. Now we show that, [$\Re\{\bu_\bk\cdot\left(\bu_\bp\times\bw_\bq\right)\}$, $\Re\{\bu_\bp\cdot\left(\bu_\bq\times\bw_\bk\right)\}$ and $\Re\{\bu_\bq\cdot\left(\bu_\bk\times\bw_\bp\right)\}$] form a set of basis for the set of all kinetic combined transfers [$S^{K}(\bk|\bp,\bq)$, $S^{K}(\bp|\bq,\bk)$ and $S^{K}(\bq|\bk,\bp)$] in a triad $(\bk,\bp,\bq)$. The completeness of the base functions is readily proved as 
\begin{align}
S^{K}(\bk|\bp,\bq)&=\Re\{\bu_\bk\cdot\left(\bu_\bp\times\bw_\bq\right)-\bu_\bq\cdot\left(\bu_\bk\times\bw_\bp\right)\},\label{SEK}\\
S^{K}(\bp|\bq,\bk)&=\Re\{\bu_\bp\cdot\left(\bu_\bq\times\bw_\bk\right)-\bu_\bk\cdot\left(\bu_\bp\times\bw_\bq\right)\},\label{SEP}\\
S^{K}(\bq|\bk,\bp)&=\Re\{\bu_\bq\cdot\left(\bu_\bk\times\bw_\bp\right)-\bu_\bp\cdot\left(\bu_\bq\times\bw_\bk\right)\}.\label{SEQ}
\end{align}
To explicitly show their independence, we have to show  
\begin{align}
A\ \Re\{\bu_\bk\cdot\left(\bu_\bp\times\bw_\bq\right)\}&+B\ \Re\{\bu_\bp\cdot\left(\bu_\bq\times\bw_\bk\right)\}\nonumber\\
&+C\ \Re\{\bu_\bq\cdot\left(\bu_\bk\times\bw_\bp\right)\}=0,
\end{align}
only if $A=B=C=0$, where $A$, $B$ and $C$ are constants. To show that, without any loss of generality, we assume $\bu_\bq=\alpha\bu_\bk=\beta\bw_\bk$ within a triad, where $\alpha$ and $\beta$ are real constants. Using these conditions in above equation one obtains, $\Re\{\bu_\bp\cdot\left(\bu_\bq\times\bw_\bk\right)\}=\Re\{\bu_\bq\cdot\left(\bu_\bk\times\bw_\bp\right)\}=0$ 
and $\Re\{\bu_\bk\cdot\left(\bu_\bp\times\bw_\bq\right)\}\neq 0$ (in general) leading to $A=0$. Similarly, for $\bu_\bk=\alpha\bu_\bp=\beta\bw_\bp$ and $\bu_\bp=\alpha\bu_\bq=\beta\bw_\bq$, we can show $B=0$ and $C=0$ respectively. Similarly, one can show that [$\Re\{\bu_\bk\cdot\left(\bj_\bp\times\bb_\bq\right)
+\bj_\bk\cdot\left(\bu_\bp\times\bb_\bq\right)\}$, 
$\Re\{\bu_\bp\cdot\left(\bj_\bq\times\bb_\bk\right)
+\bj_\bp\cdot\left(\bu_\bq\times\bb_\bk\right)\}$ and $\Re\{\bu_\bq\cdot\left(\bj_\bk\times\bb_\bp\right)
+\bj_\bq\cdot\left(\bu_\bk\times\bb_\bp\right)\}$] form a basis for [$S^{M}(\bk|\bp,\bq)$, $S^{M}(\bp|\bq,\bk)$ and $S^{M}(\bq|\bk,\bp)$] and their completeness can be shown in a way similar to Eqs. \eqref{SEK}-\eqref{SEQ} and their independence can shown by consecutively choosing (i) $\bb_\bp=\alpha \bu_\bq=\beta\bj_\bq$ along with $\bb_\bk=\alpha \bu_\bp=\beta\bj_\bp$, (ii) $\bb_\bq=\alpha \bu_\bk=\beta\bj_\bk$ along with $\bb_\bp=\alpha \bu_\bq=\beta\bj_\bq$ and (iii) $\bb_\bq=\alpha \bu_\bk=\beta\bj_\bk$ along with $\bb_\bk=\alpha \bu_\bp=\beta\bj_\bp$. Finally for the Hall term one can show that, [$\Re\{\bj_\bk\cdot(\bb_\bq\times\bj_\bp)\}$, $\Re\{\bj_\bp\cdot(\bb_\bk\times\bj_\bq)\}$ and
$\Re\{\bj_\bq\cdot(\bb_\bp\times\bj_\bk)\}$] constitute a set of basis functions for [$S^{\pazocal{H}}(\bk|\bp,\bq)$, $S^{\pazocal{H}}(\bp|\bq,\bk)$ and $S^{\pazocal{H}}(\bq|\bk,\bp)$] with a similar proof of completeness shown in Eqs.~\eqref{SEK}-\eqref{SEQ} and the independence can be obtained with the consecutive choices (i) $\bb_\bp=\alpha\bj_\bq=\beta\bj_\bp$, (ii) $\bb_\bq=\alpha\bj_\bk=\beta\bj_\bq$ and (iii) $\bb_\bk=\alpha\bj_\bp=\beta\bj_\bk$. 

Note that, the aforementioned choices assumed 
$\bu$-$\bw$, $\bu$-$\bj$ and $\bb$-$\bj$ alignments for certain modes in a triad. However, such alignments are not true for all the three modes in a triad and hence does not correspond to the alignment at each point in real space. The above analysis clearly shows that each of the combined transfers rates $S^K,\ S^{M}$ and $S^{\pazocal{H}}$ can be expressed as a linear combination of three base functions. In case of a Beltrami state, the aligned condition is true for every mode in a triad and all the combined transfer rates identically vanish. However, the M2M transfer rates still may survive with $S(\bk|\bp|\bq)=-S(\bk|\bq|\bp)\neq 0$ thus characterising the aligned state. As mentioned previously, due to the particular structure of Eqs.~\eqref{SEK}-\eqref{SEQ}, the choice of base functions is not unique and an equivalent basis can always be obtained if all the base functions are added to an arbitrary circulation function $\Delta$. Nevertheless, unlike \citet{Plunian_2019}, here we can meaningfully determine the dimension of the basis corresponding to the combined transfer rates which are uniquely defined. By construction, one can effectively define a single combined transfer rate $S_T=S^K+S^M+d_i S^{\pazocal{H}}$ with the base functions [$S_T (\bk|\bp|\bq)$, $S_T (\bp|\bq|\bk)$ and $S_T (\bq|\bk|\bp)$] where
\begin{equation}
S_T (\bk|\bp|\bq)=S^K(\bk|\bp|\bq)+S^M(\bk|\bp|\bq)+d_i S^{\pazocal{H}}(\bk|\bp|\bq).
\end{equation}
Proceeding similar to Eqs.~\eqref{SEK}-\eqref{SEQ}, one can show the completeness whereas the independence is obtained by successively choosing, (i) $\bw_\bk=\alpha\bu_\bk=\beta\bu_\bq=\gamma\bb_\bp=\delta\bj_\bq=\kappa\bb_\bk$, (ii) $\bw_\bp=\alpha\bu_\bp=\beta\bu_\bk=\gamma\bb_\bq=\delta\bj_\bk=\kappa\bb_\bp$ and (iii) $\bw_\bq=\alpha\bu_\bq=\beta\bu_\bp=\gamma\bb_\bk=\delta\bj_\bp=\kappa\bb_\bq$. By definition, $S_T (\bk|\bp|\bq)$ can be thought to be the M2M transfer rate including all types of interactions owing to the energy transfer and can be called a `mode-specific' M2M transfer.

Mode-specific M2M transfer rates are necessary for obtaining the total energy transfer rate from one mode to the other inside the inertial zone. Such a knowledge is also important for the understanding of turbulent relaxation. According to the recently proposed\citep{Banerjee_2023} principle of vanishing nonlinear transfer (PVNLT), a turbulent relaxed state is characterized by a halt in the nonlinear energy cascade which, in turn, leads to the vanishing of the total mode-specific combined transfer rates in the inertial zone. One thus expects 
\begin{align}
&\int\limits_{\bk + \bp + \bq= {\bf 0}}
\left[\bu_\bk\cdot(\bu_\bp\times\bw_\bq+\bu_\bq\times\bw_\bp+\bj_\bp\times\bb_\bq+\bj_\bq\times\bb_\bp)\right.\nonumber\\
&\left.+\bj_\bk\cdot \{(\bu_\bp-d_i\bj_\bp)\times\bb_\bq+(\bu_\bq-d_i\bj_\bq)\times\bb_\bp\}  \right] d {\bp} =0
\end{align}
at all scales inside the inertial zone. In the most general case, for non-zero $\bu_\bk$ and $\bj_\bk$, the above equation trivially vanishes if 
\begin{align}
\bu_\bp\times\bw_\bq+\bu_\bq\times\bw_\bp+\bj_\bp\times\bb_\bq+\bj_\bq\times\bb_\bp&=i\bk\phi_k,\\   
(\bu_\bp-d_i\bj_\bp)\times\bb_\bq+(\bu_\bq-d_i\bj_\bq)\times\bb_\bp&=i\bk\psi_k
\end{align}
where $\phi_\bk$ and $\psi_\bk$ are arbitrary scalar functions and both $\bk\cdot\bu_\bk$ and $\bk\cdot\bj_\bk$ are zero in incompressible HMHD. In the real space, the corresponding relaxed states are given by the Fourier transforms of the aforesaid equations as 
\begin{equation}
    \bu\times\bw+\bj\times\bb =\bnabla\phi,\;\; \text{and}\;\;(\bu-d_i\bj)\times\bb =\bnabla\psi.
\end{equation}
The obtained relaxed states are exactly identical to those obtained in Eqs.~(26) and (27) of Ref.\citep{Banerjee_2023}.

\section{field-specific M2M transfer rates}
Despite aforementioned importance, mode-specific M2M transfers do not allow us to probe into individual contributions of diverse interactions between modal field variables $\bu_\bk$, $\bb_\bk$ etc. in the cascade of energy and one therefore has to investigate the field-specific M2M transfers.

In ordinary MHD, there are four types of possible interactions- (i) $\bu$-to-$\bu$ ($S^{uu}$), (ii) $\bb$-to-$\bu$ ($S^{ub}$), (iii) $\bu$-to-$\bb$ ($S^{bu}$) and (iv) $\bb$-to-$\bb$ ($S^{bb}$). By careful observation, one can in fact recognise that
\begin{align}
S^K\left(\bk|\bp,\bq\right)&=S^{uu}\left(\bk|\bp,\bq\right),\label{mode2field_K}\\
S^M\left(\bk|\bp,\bq\right)&=S^{ub}\left(\bk|\bp,\bq\right)+S^{bu}\left(\bk|\bp,\bq\right)+S^{bb}\left(\bk|\bp,\bq\right),\label{mode2field_M}
\end{align}
where
\begin{align}
S^{uu}(\bk|\bp,\bq)&=\Re[i\bu_\bk\cdot\{\bk\cdot(\bu_\bq\otimes\bu_\bp)\}+i\bu_\bk\cdot\{\bk\cdot(\bu_\bp\otimes\bu_\bq)\}]\nonumber\\
&=\Re\{i(\bk\cdot\bu_\bq)(\bu_\bp\cdot\bu_\bk)+i(\bk\cdot\bu_\bp)(\bu_\bq\cdot\bu_\bk)\},\label{Suu_combined}\\
S^{ub}(\bk|\bp,\bq)&=\Re[-i\bu_\bk\cdot\{\bk\cdot(\bb_\bq\otimes\bb_\bp)\}-i\bu_\bk\cdot\{\bk\cdot(\bb_\bp\otimes\bb_\bq)\}]\nonumber\\
&=\Re\{-i(\bk\cdot\bb_\bq)(\bb_\bp\cdot\bu_\bk)-i(\bk\cdot\bb_\bp)(\bb_\bq\cdot\bu_\bk)\},\label{Sub_combined}\\
S^{bu}(\bk|\bp,\bq)&=\Re[-i\bb_\bk\cdot\{\bk\cdot(\bb_\bq\otimes\bu_\bp)\}-i\bb_\bk\cdot\{\bk\cdot(\bb_\bp\otimes\bu_\bq)\}]\nonumber\\
&=\Re\{-i(\bk\cdot\bb_\bq)(\bu_\bp\cdot\bb_\bk)-i(\bk\cdot\bb_\bp)(\bu_\bq\cdot\bb_\bk)\},\label{Sbu_combined}\\
S^{bb}(\bk|\bp,\bq)&=\Re[i\bb_\bk\cdot\{\bk\cdot(\bu_\bq\otimes\bb_\bp)\}+i\bb_\bk\cdot\{\bk\cdot(\bu_\bp\otimes\bb_\bq)\}]\nonumber\\
&=\Re\{i(\bk\cdot\bu_\bq)(\bb_\bp\cdot\bb_\bk)+i(\bk\cdot\bu_\bp)(\bb_\bq\cdot\bb_\bk)\}.\label{Sbb_combined}
\end{align}
At this point we deliberately omit the similar expression for the Hall term $S^{\pazocal{H}}\left(\bk|\bp,\bq\right)$ which will be discussed later. From Eqs.~\eqref{Suu_combined}-\eqref{Sbb_combined}, it is easy to see that $S^{uu}$ and $S^{bb}$ individually follow a triadic conservation whereas $S^{ub}$ and $S^{bu}$ together satisfy a triadic conservation. In accordance with the definitions 
\begin{align}
S^{xy}(\bk|\bp|\bq)&=-S^{yx}(\bp|\bk|\bq)\label{anti_field_specific}\;\;\;\text{and}\\ 
S^{xy}(\bk|\bp,\bq)&=S^{xy}(\bk|\bp|\bq)+S^{xy}(\bk|\bq|\bp),
\end{align}
it is rather straightforward to write (omitting the circulation transfers) the corresponding M2M transfers rates as\citep{DAR_2001, Verma_2004, Plunian_2019}
\begin{align}
&S^{uu}(\bk|\bp|\bq)\nonumber\\
&=\Re[i\bu_\bk\cdot\{\bk\cdot(\bu_\bq\otimes\bu_\bp)\}]=\Re\{i(\bk\cdot\bu_\bq)(\bu_\bp\cdot\bu_\bk)\},\label{Suu}\\
&S^{ub}(\bk|\bp|\bq)\nonumber\\
&=\Re[-i\bu_\bk\cdot\{\bk\cdot(\bb_\bq\otimes\bb_\bp)\}]=\Re\{-i(\bk\cdot\bb_\bq)(\bb_\bp\cdot\bu_\bk))\},\label{Sub}\\
&S^{bu}(\bk|\bp|\bq)\nonumber\\
&=\Re[-i\bb_\bk\cdot\{\bk\cdot(\bb_\bq\otimes\bu_\bp)\}]=\Re\{-i(\bk\cdot\bb_\bq)(\bu_\bp\cdot\bb_\bk))\},\label{Sbu}\\
&S^{bb}(\bk|\bp|\bq)\nonumber\\
&=\Re[i\bb_\bk\cdot\{\bk\cdot(\bu_\bq\otimes\bb_\bp)\}]=\Re\{i(\bk\cdot\bu_\bq)(\bb_\bp\cdot\bb_\bk))\}.\label{Sbb}
\end{align}
In contrast with mode-specific M2M transfer rates, here both the tensorial and vectorial formulation give us an opportunity to unambiguously determine the giver, receiver and mediator fields. For example, the generic M2M transfer $S^{xy}(\bk|\bp|\bq)=\Re[i\bx_\bk\cdot\{\bk\cdot(\bz_\bq\otimes\by_\bp)\}]=\Re\{i(\bk\cdot\bz_\bq)(\by_\bp\cdot\bx_\bk)\}$ can be associated to the transfer rate of energy from $\by_\bp$ to $\bx_\bk$ with $\bz_\bq$ as a mediator. It is therefore reasonable to call $S^{xy}(\bk|\bp|\bq)$ a `field-specific' M2M transfer rate. Contrary to Eq.~\eqref{m2mK}, here one needs to explicitly use the incompressibility condition to satisfy the giver-receiver antisymmetry in Eq.~\eqref{anti_field_specific} as
\begin{align}
&S^{xy}(\bk|\bp|\bq)+S^{yx}(\bp|\bk|\bq)\nonumber\\
&=\Re\{i(\bk\cdot\bz_\bq)(\by_\bp\cdot\bx_\bk)\}+\Re\{i(\bp\cdot\bz_\bq)(\bx_\bk\cdot\by_\bp)\}\\
&=\Re\{i((\bk+\bp)\cdot\bz_\bq)(\by_\bp\cdot\bx_\bk)\}=-\Re\{i(\bq\cdot\bz_\bq)(\by_\bp\cdot\bx_\bk)\}=0.\nonumber
\end{align}
This is simply because the tensor formulation, unlike the previous formulation, lacks the giver-receiver antisymmetry.  

A natural question that comes up at this point is whether the above mentioned field-specific M2M transfer rates also constitute a basis for the corresponding combined transfer rates. For $S^{uu}$, we need to show 
\begin{align}
    A\ \Re\{i(\bk\cdot\bu_\bq)(\bu_\bp\cdot\bu_\bk)\}&+B\ \Re\{i(\bq\cdot\bu_\bp)(\bu_\bk\cdot\bu_\bq)\}\nonumber\\
    &+C\ \Re\{i(\bp\cdot\bu_\bk)(\bu_\bq\cdot\bu_\bp)\}=0,\label{basis_uu_field_specific}
\end{align}
only if $A=B=C=0$. The completeness of the basis can be easily shown using the expression for combined transfer rates following Eq.~\eqref{Suu_combined} and their independence can be shown by assuming, without losing generality, that $\bu_\bk\cdot\bu_\bq=\bu_\bq\cdot\bu_\bp=0$. Using this fact in Eq.~\eqref{basis_uu_field_specific}, one gets $\Re\{i(\bq\cdot\bu_\bp)(\bu_\bk\cdot\bu_\bq)\}=\Re\{i(\bp\cdot\bu_\bk)(\bu_\bq\cdot\bu_\bp)\}=0$ and $\Re\{i(\bk\cdot\bu_\bq)(\bu_\bp\cdot\bu_\bk)\}\neq 0$ leading to $A=0$. Similarly one can show $B=0$ and $C=0$. Again for $S^{bb}$, one can show that [$\Re\{i(\bk\cdot\bu_\bq)(\bb_\bp\cdot\bb_\bk)\}$, $\Re\{i(\bq\cdot\bu_\bp)(\bb_\bk\cdot\bb_\bq)\}$ and $\Re\{i(\bp\cdot\bu_\bk)(\bb_\bq\cdot\bb_\bp)\}$] form a basis. The completeness can be shown using Eq.~\eqref{Sbb_combined} and to prove their independence one assumes the pairwise vanishing of $(\bb_\bk\cdot\bb_\bq)$, $(\bb_\bp\cdot\bb_\bq)$ and $(\bb_\bk\cdot\bb_\bp)$ respectively. Since $S^{ub}$ and $S^{bu}$ together satisfy a triadic conservation, it is reasonable to define a basis for the combined transfer $S^{ub}+S^{bu}$. One can indeed show that [$\Re\{i(\bk\cdot\bb_\bq)(\bb_\bp\cdot\bu_\bk+\bu_\bp\cdot\bb_\bk)\}$, $\Re\{i(\bq\cdot\bb_\bp)(\bb_\bk\cdot\bu_\bq+\bu_\bk\cdot\bb_\bq)\}$ and $\Re\{i(\bp\cdot\bb_\bk)(\bb_\bq\cdot\bu_\bp+\bu_\bq\cdot\bb_\bp)\}$] form a basis for $S^{ub}+S^{bu}$. The completeness of the basis is proved following Eqs. \eqref{Sub_combined} and \eqref{Sbu_combined} and their independence is obtained by consecutively assuming (i) $\bu_\bp=\alpha\bu_\bk$ and $\bb_\bp=\beta\bb_\bk$ along with $\bb_\bq\cdot\bu_\bp=0$ and $\bb_\bp\cdot\bu_\bq=0$, (ii) $\bu_\bq=\alpha\bu_\bk$ and $\bb_\bq=\beta\bb_\bk$ along with $\bu_\bq\cdot\bb_\bp=0$ and $\bu_\bp\cdot\bb_\bk=0$ and (iii) $\bu_\bp=\alpha\bu_\bq$ and $\bb_\bp=\beta\bb_\bq$ along with $\bu_\bp\cdot\bb_\bk=0$ and $\bb_\bp\cdot\bu_\bk=0$.


\section{fundamental units of nonlinear transfers due to the Hall term}
Unlike mode-specific transfer rates, obtaining field-specific transfer rates for the Hall term is tricky. Several previous studies did not distinguish between $\bb$ and $\bj$-fields and the Hall transfer was simply interpreted as a transfer between different modes of the $\bb$-field\citep{Mininni_2007,Gomez_2010,Miura_2019}. However, $\bb$ and $\bj$-field are independent and in Fourier space they are related as $\bj_\bk=i\bk\times\bb_\bk$\citep{Halder_2023, Meyrand_2018}. It is thus logical to decompose the Hall term as $-\bnabla\times\left(\bj\times\bb\right)=-\left(\bb\cdot\bnabla\right)\bj+\left(\bj\cdot\bnabla\right)\bb$. In the magnetic energy transfer rate, the first term on the right is responsible for $\bj$-to-$\bb$ transfer and the latter is responsible for $\bb$-to-$\bb$ transfer. The role of such a decomposition in HMHD dynamo action has been studied recently, where the small-scale current fields are found to be the primary contributors to the generation of large-scale magnetic fields \citep{Halder_2023}. By construction, the Hall term does not include the back transfer corresponding to the $\bj$-to-$\bb$ transfer. Following Eqs.~\eqref{mode2field_K} and \eqref{mode2field_M}, one can write 
\begin{equation}
S^{\pazocal{H}}\left(\bk|\bp,\bq\right)=\Sigma^{bj}\left(\bk|\bp,\bq\right)+\Sigma^{bb}\left(\bk|\bp,\bq\right),\label{mode2field_H} \end{equation}
where
\begin{align}
\Sigma^{bj}(\bk|\bp,\bq)&=\Re[i\bb_\bk\cdot\{\bk\cdot(\bb_\bq\otimes\bj_\bp)\}+i\bb_\bk\cdot\{\bk\cdot(\bb_\bp\otimes\bj_\bq)\}]\nonumber\\
&=\Re\{i(\bk\cdot\bb_\bq)(\bj_\bp\cdot\bb_\bk)+i(\bk\cdot\bb_\bp)(\bj_\bq\cdot\bb_\bk)\},\label{Sbj_H_combined}\\
\Sigma^{bb}(\bk|\bp,\bq)&=\Re[-i\bb_\bk\cdot\{\bk\cdot(\bj_\bq\otimes\bb_\bp)\}-i\bb_\bk\cdot\{\bk\cdot(\bj_\bp\otimes\bb_\bq)\}]\nonumber\\
&=\Re\{-i(\bk\cdot\bj_\bq)(\bb_\bp\cdot\bb_\bk)-i(\bk\cdot\bj_\bp)(\bb_\bq\cdot\bb_\bk)\}.\label{Sbb_H_combined}
\end{align}
Similar to $S^{bb}$ above, one can show that [$\Re\{i(\bk\cdot\bj_\bq)(\bb_\bp\cdot\bb_\bk)\}$, $\Re\{i(\bq\cdot\bj_\bp)(\bb_\bk\cdot\bb_\bq)\}$ and $\Re\{i(\bp\cdot\bj_\bk)(\bb_\bq\cdot\bb_\bp)\}$] form a basis for $\Sigma^{bb}$ and their completeness and independence can also be proven similarly. For $\Sigma^{bj}$, obtaining such a basis is not straightforward as the Hall term does not allow $\Sigma^{jb}$ transfers thereby making condition \eqref{anti_field_specific} invalid. Following Eq.~\eqref{Sbj_H_combined}, the combined transfer rate for the $\bp$ and $\bq$-th mode can be written as 
\begin{align}
\Sigma^{bj}(\bp|\bq,\bk)
&=\Re\{i(\bp\cdot\bb_\bq)(\bj_\bk\cdot\bb_\bp)+i(\bp\cdot\bb_\bk)(\bj_\bq\cdot\bb_\bp)\},\label{Sbj_H_combined-p}\\
\Sigma^{bj}(\bq|\bk,\bp)
&=\Re\{i(\bq\cdot\bb_\bk)(\bj_\bp\cdot\bb_\bq)+i(\bq\cdot\bb_\bp)(\bj_\bk\cdot\bb_\bq)\}.\label{Sbj_H_combined-q}
\end{align}
In contrast to the previous cases, due to the lack of giver-receiver antisymmetry, here one cannot simply define a three member basis. Instead, using the triadic conservation property of $\Sigma^{bj}$, one can define a five component basis consisting of any five of six M2M transfer rates present in Eqs.~\eqref{Sbj_H_combined}, \eqref{Sbj_H_combined-p} and \eqref{Sbj_H_combined-q}.
\begin{figure}[h!]
    \centering
    \includegraphics[width=0.5\linewidth]{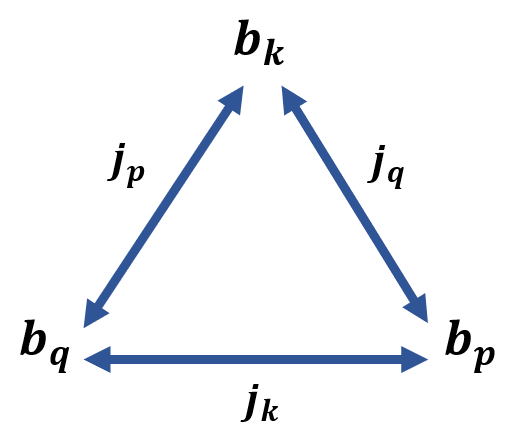}
    \caption{Fundamental units of transfer in $\Sigma^{bb}$.}
    \label{fig:Hall_bb_pheno}
\end{figure}
\begin{figure}[h!]
    \centering
    \includegraphics[width=0.5\linewidth]{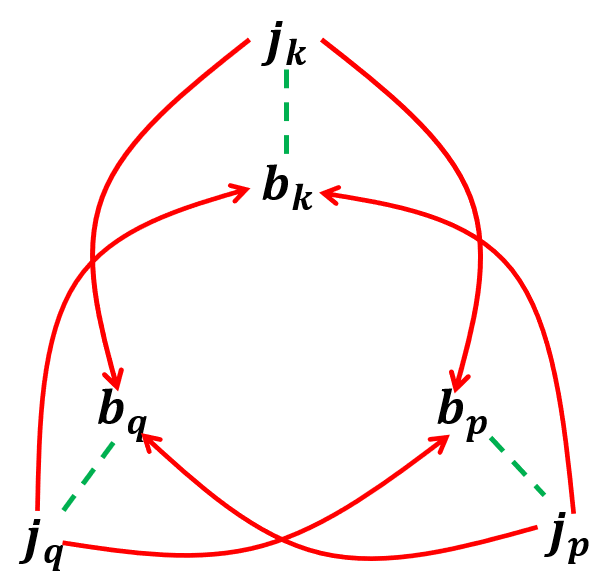}
    \caption{Fundamental units of transfer in $\Sigma^{bj}$.}
    \label{fig:Hall_jb_pheno}
\end{figure}
To that end, we chose [$\Re\{i(\bk\cdot\bb_\bq)(\bj_\bp\cdot\bb_\bk)\}$, $\Re\{i(\bk\cdot\bb_\bp)(\bj_\bq\cdot\bb_\bk)\}$, $\Re\{i(\bp\cdot\bb_\bq)(\bj_\bk\cdot\bb_\bp)\}$, $\Re\{i(\bp\cdot\bb_\bk)(\bj_\bq\cdot\bb_\bp)\}$ and $\Re\{i(\bq\cdot\bb_\bk)(\bj_\bp\cdot\bb_\bq)\}$] as our basis. The completeness of such a basis is again proved using Eqs.~\eqref{Sbj_H_combined}, \eqref{Sbj_H_combined-p} and \eqref{Sbj_H_combined-q} and the independence is proven by consecutively choosing (i) $\bb_\bk\cdot\bj_\bq=\bb_\bp\cdot\bj_\bk=\bb_\bp\cdot\bj_\bq=\bb_\bq\cdot\bj_\bp=0$, (ii) $\bb_\bk\cdot\bj_\bp=\bb_\bp\cdot\bj_\bk=\bb_\bp\cdot\bj_\bq=\bb_\bq\cdot\bj_\bp=0$, (iii) $\bb_\bk\cdot\bj_\bp=\bb_\bk\cdot\bj_\bq=\bb_\bp\cdot\bj_\bq=\bb_\bq\cdot\bj_\bp=0$, (iv) $\bb_\bk\cdot\bj_\bp=\bb_\bk\cdot\bj_\bq=\bb_\bp\cdot\bj_\bk=\bb_\bq\cdot\bj_\bp=0$ and (v) $\bb_\bk\cdot\bj_\bp=\bb_\bk\cdot\bj_\bq=\bb_\bp\cdot\bj_\bk=\bb_\bp\cdot\bj_\bq=0$. 

The dimensionality of the basis for $\Sigma^{bb}$ and $\Sigma^{bj}$ can be understood by means of simple  figures provided in Figs.~\ref{fig:Hall_bb_pheno} and \ref{fig:Hall_jb_pheno}. The three blue bi-directional arrows in Fig.~\ref{fig:Hall_bb_pheno} constitute six possible M2M transfer rates for $\Sigma^{bb}$. This reduces to three due to the giver-receiver antisymmetry. Switching off any two of the three aforementioned transfers does not imply the vanishing of the third thus justifying the independence of the three fundamental (non-unique) transfers of $\Sigma^{bb}$. For $\Sigma^{bj}$, one can similarly find six possible transfer rates (denoted by red uni-directional arrows in Fig.~\ref{fig:Hall_jb_pheno}). Although independent, the modification of $\bb_\bk$ also modifies $\bj_\bk$ and vice versa. The green dashed lines in Fig.~\ref{fig:Hall_jb_pheno} denote such modifications and not the nonlinear mode-to-mode transfers. Unlike $\Sigma^{bb}$, these six transfers do not reduce to three due to the absence of $\bb$-to-$\bj$ back transfers. However, the sum of these six transfers vanishes satisfying a triadic conservation (see appendix \ref{appA}). If any five of them are independently chosen to be zero, then remaining one identically vanishes due to the detailed triadic conservation indicating that not all six of them are independent. Similarly, if any four of them are switched off then the sum of the remaining two becomes zero making them equal and opposite to each other. Both of these facts indicate that five independent transfer rates are necessary to constitute $\Sigma^{bj}$ interactions.

Although the fundamental units of interaction for $\Sigma^{bj}$ do not obey the giver-receiver anti-symmetry, they also possess an inherent non-uniqueness. For example, if we define a new set of fundamental units as [$\Re\{i(\bk\cdot\bb_\bq)(\bj_\bp\cdot\bb_\bk)\}+\Delta_1$, $\Re\{i(\bk\cdot\bb_\bp)(\bj_\bq\cdot\bb_\bk)\}-\Delta_1$, $\Re\{i(\bp\cdot\bb_\bq)(\bj_\bk\cdot\bb_\bp)\}+\Delta_2$, $\Re\{i(\bp\cdot\bb_\bk)(\bj_\bq\cdot\bb_\bp)\}-\Delta_2$ and $\Re\{i(\bq\cdot\bb_\bk)(\bj_\bp\cdot\bb_\bq)\}+\Delta_3$, $\Re\{i(\bq\cdot\bb_\bp)(\bj_\bk\cdot\bb_\bq)\}-\Delta_3$], then the combined transfer rates in Eqs. \eqref{Sbj_H_combined}, \eqref{Sbj_H_combined-p} and \eqref{Sbj_H_combined-q} as well as the triadic conservation remain unchanged. The nature of non-uniqueness is visibly different than that for M2M transfers. However, in the force-free situation, where we have a $\bj$-$\bb$ alignment for every mode in the triad \textit{i.e.} $\bj_\bk=\alpha\bb_\bk$, $\bj_\bp=\alpha\bb_\bp$ and $\bj_\bq=\alpha\bb_\bq$, the $\bj$-$\bb$ transfers practically become $\bb$-$\bb$ transfer and the corresponding non-uniqueness becomes of Kraichnan type. 

\section{Discussion}
In this work, we have suggested a rigorous framework to search for the fundamental units of triad interaction in fully developed turbulence. While the combined transfers can always be uniquely determined, the M2M transfers are non-unique by definition. Here, we have investigated the fundamental units of the combined transfers in HMHD turbulence. This current work provides a straightforward rectification over the approach provided in Ref.\citep{Plunian_2019}, where the basis elements for non-unique M2M transfers were investigated. Further as it is shown here, the fundamental units of triadic interactions can be constructed in both mode-specific and field-specific ways which have their own importance. Whereas the mode-specific transfer rates are pivotal to calculate the scale-specific contribution to the energy cascade and to characterize the turbulent relaxed states, the field-specific transfer rates become particularly useful for calculating and comparing the individual contributions of the interacting modal fields in the study of energy cascade and turbulent dynamos\citep{Kumar_2013, Kumar_2017, Halder_2023}. For HMHD turbulence, we have obtained the basis for all the combined transfers in a triad by proving their completeness and independence. In mode-specific case, all the combined transfers are found to be constituted by corresponding M2M transfers which are inherently non-unique. For the field-specific case, all the combined transfers of MHD consist of non-unique M2M transfers. The Hall contribution can be decomposed into $\bb$-$\bb$ and $\bj$-$\bb$ interactions. Whereas the first one consists of non-unique M2M transfers and is associated with a three-member basis, the second one does not contain M2M transfers due to the absence of $\bb$-$\bj$ back transfer and the corresponding combined transfers can be constituted by a five-member basis where the fundamental units of transfer are associated with a different kind of non-uniqueness which does not come out of giver-receiver anti-symmetry. Despite the inherent non-uniqueness of the fundamental units of transfers, a precise knowledge of their number is crucial both for phenomenological understanding and computational purpose, especially for making the numerical schemes more cost effective. Our formalism can be used for more complex turbulent fluid flows \textit{e.g.} ferrofluids, binary fluids etc, where universal energy cascades are also studied\citep{Mouraya_2019, Pan_2022}.   


\section*{author declarations}
\subsection*{Conflict of interest}
The authors declare no conflicts of interest.
\subsection*{Data availability} 
Not applicable as our work is purely analytical and does not contain any data.
\section*{author contributions}
\noindent\textbf{Supratik Banerjee:} Conceptualization (lead); Formal analysis (lead); Writing - original draft (supporting).  \\
\textbf{Arijit Halder:} Conceptualization (supporting); Formal analysis (supporting); Writing - original draft (lead).

\section{Appendixes}

\appendix
\section{Triadic conservation of $\Sigma^{bj}$}
\label{appA}

From Eqs.~\eqref{Sbj_H_combined}, \eqref{Sbj_H_combined-p} and \eqref{Sbj_H_combined-q}, the combined transfer rates are given by
\begin{align}
   \Sigma^{bj}(\bk|\bp,\bq)&=\Re\{i(\bk\cdot\bb_\bq)(\bj_\bp\cdot\bb_\bk)+i(\bk\cdot\bb_\bp)(\bj_\bq\cdot\bb_\bk)\},\\
   \Sigma^{bj}(\bp|\bq,\bk)&=\Re\{i(\bp\cdot\bb_\bq)(\bj_\bk\cdot\bb_\bp)+i(\bp\cdot\bb_\bk)(\bj_\bq\cdot\bb_\bp)\},\\
   \Sigma^{bj}(\bq|\bk,\bp)&=\Re\{i(\bq\cdot\bb_\bp)(\bj_\bk\cdot\bb_\bq)+i(\bq\cdot\bb_\bk)(\bj_\bp\cdot\bb_\bq)\}.
\end{align}
Using the relations $\bj_\bk=i\bk\times\bb_\bk$, $\bj_\bp=i\bp\times\bb_\bp$ and $\bj_\bq=i\bq\times\bb_\bq$ in the above equations and rearranging one gets 
\begin{align}
   &\Sigma^{bj}(\bk|\bp,\bq)\nonumber\\
   &=\Re[(\bk\cdot\bb_\bq)\{\bp\cdot(\bb_\bk\times\bb_\bp)\}+(\bk\cdot\bb_\bp)\{\bq\cdot(\bb_\bk\times\bb_\bq)\}],\label{S_Hall_jb_k}\\
   &\Sigma^{bj}(\bp|\bq,\bk)\nonumber\\
   &=\Re[(\bp\cdot\bb_\bq)\{\bk\cdot(\bb_\bp\times\bb_\bk)\}+(\bp\cdot\bb_\bk)\{\bq\cdot(\bb_\bp\times\bb_\bq)\}],\label{S_Hall_jb_p}\\
   &\Sigma^{bj}(\bq|\bk,\bp)\nonumber\\
   &=\Re[(\bq\cdot\bb_\bp)\{\bk\cdot(\bb_\bq\times\bb_\bk)\}+(\bq\cdot\bb_\bk)\{\bp\cdot(\bb_\bq\times\bb_\bp)\}].\label{S_Hall_jb_q}
\end{align}
Adding Eqs.~\eqref{S_Hall_jb_k}-\eqref{S_Hall_jb_q} and using imcompressibilty condition, one obtains
\begin{align}
\Sigma^{bj}(\bk|\bp,\bq)&+\Sigma^{bj}(\bp|\bq,\bk)+\Sigma^{bj}(\bq|\bk,\bp) \nonumber\\
=&\Re[(\bk\cdot\bb_\bq)\{\bp\cdot(\bb_\bk\times\bb_\bp)-\bk\cdot(\bb_\bp\times\bb_\bk)\}\nonumber\\
&+(\bq\cdot\bb_\bp)\{\bk\cdot(\bb_\bq\times\bb_\bk)-\bq\cdot(\bb_\bk\times\bb_\bq)  \}\nonumber\\
&+(\bp\cdot\bb_\bk)\{\bq\cdot(\bb_\bp\times\bb_\bq)-\bp\cdot(\bb_\bq\times\bb_\bp) \}].\label{Sumtrans1}
\end{align}
Using $(\bf A\times\bf B)=-(\bf B\times\bf A)$ and the triadic constraint $\bk+\bp+\bq=\bf 0$, one can write  
\begin{align}
&\Sigma^{bj}(\bk|\bp,\bq)+\Sigma^{bj}(\bp|\bq,\bk)+\Sigma^{bj}(\bq|\bk,\bp) \nonumber\\
&=\Re[(\bk\cdot\bb_\bq)\{\bq\cdot(\bb_\bp\times\bb_\bk)\}+(\bq\cdot\bb_\bp)\{\bp\cdot(\bb_\bk\times\bb_\bq)\}\nonumber\\
&+(\bp\cdot\bb_\bk)\{\bk\cdot(\bb_\bq\times\bb_\bp)\} ]. \label{Sumtrans3}
\end{align}
Using the relation $(\bf A\times\bf B)\cdot(\bf C\times\bf D)=(\bf A\cdot\bf C)(\bf B\cdot\bf D)-(\bf B\cdot\bf C)(\bf A\cdot\bf D)$, the above equation reduces to
\begin{align}
&\Sigma^{bj}(\bk|\bp,\bq)+\Sigma^{bj}(\bp|\bq,\bk)+\Sigma^{bj}(\bq|\bk,\bp) \\
&=\Re[(\bk\times\bq)\cdot\{\bb_\bq\times(\bb_\bp\times\bb_\bk)+\bb_\bk\times(\bb_\bq\times\bb_\bp)+\bb_\bp\times(\bb_\bk\times\bb_\bq)\}].\nonumber
\end{align}
The \textit{r.h.s.} of the above equation vanishes due to the Jacobi identity, leading to  
\begin{equation}
    \Sigma^{bj}(\bk|\bp,\bq)+\Sigma^{bj}(\bp|\bq,\bk)+\Sigma^{bj}(\bq|\bk,\bp)=0.
\end{equation}

%


\begin{thebibliography}{37}%
\makeatletter
\providecommand \@ifxundefined [1]{%
 \@ifx{#1\undefined}
}%
\providecommand \@ifnum [1]{%
 \ifnum #1\expandafter \@firstoftwo
 \else \expandafter \@secondoftwo
 \fi
}%
\providecommand \@ifx [1]{%
 \ifx #1\expandafter \@firstoftwo
 \else \expandafter \@secondoftwo
 \fi
}%
\providecommand \natexlab [1]{#1}%
\providecommand \enquote  [1]{``#1''}%
\providecommand \bibnamefont  [1]{#1}%
\providecommand \bibfnamefont [1]{#1}%
\providecommand \citenamefont [1]{#1}%
\providecommand \href@noop [0]{\@secondoftwo}%
\providecommand \href [0]{\begingroup \@sanitize@url \@href}%
\providecommand \@href[1]{\@@startlink{#1}\@@href}%
\providecommand \@@href[1]{\endgroup#1\@@endlink}%
\providecommand \@sanitize@url [0]{\catcode `\\12\catcode `\$12\catcode
  `\&12\catcode `\#12\catcode `\^12\catcode `\_12\catcode `\%12\relax}%
\providecommand \@@startlink[1]{}%
\providecommand \@@endlink[0]{}%
\providecommand \url  [0]{\begingroup\@sanitize@url \@url }%
\providecommand \@url [1]{\endgroup\@href {#1}{\urlprefix }}%
\providecommand \urlprefix  [0]{URL }%
\providecommand \Eprint [0]{\href }%
\providecommand \doibase [0]{http://dx.doi.org/}%
\providecommand \selectlanguage [0]{\@gobble}%
\providecommand \bibinfo  [0]{\@secondoftwo}%
\providecommand \bibfield  [0]{\@secondoftwo}%
\providecommand \translation [1]{[#1]}%
\providecommand \BibitemOpen [0]{}%
\providecommand \bibitemStop [0]{}%
\providecommand \bibitemNoStop [0]{.\EOS\space}%
\providecommand \EOS [0]{\spacefactor3000\relax}%
\providecommand \BibitemShut  [1]{\csname bibitem#1\endcsname}%
\let\auto@bib@innerbib\@empty
\bibitem [{\citenamefont {Frisch}\ and\ \citenamefont
  {Kolmogorov}(1995)}]{Frisch_1995}%
  \BibitemOpen
  \bibfield  {author} {\bibinfo {author} {\bibfnamefont {U.}~\bibnamefont
  {Frisch}}\ and\ \bibinfo {author} {\bibfnamefont {A.}~\bibnamefont
  {Kolmogorov}},\ }\href {https://books.google.co.in/books?id=K-Pf7RuYkf0C}
  {\emph {\bibinfo {title} {Turbulence: The Legacy of A. N. Kolmogorov}}}\
  (\bibinfo  {publisher} {Cambridge University Press},\ \bibinfo {year}
  {1995})\BibitemShut {NoStop}%
\bibitem [{\citenamefont {Tennekes}\ and\ \citenamefont
  {Lumley}(2018)}]{Tennekes_2018}%
  \BibitemOpen
  \bibfield  {author} {\bibinfo {author} {\bibfnamefont {H.}~\bibnamefont
  {Tennekes}}\ and\ \bibinfo {author} {\bibfnamefont {J.~L.}\ \bibnamefont
  {Lumley}},\ }\href@noop {} {\emph {\bibinfo {title} {A first course in
  turbulence}}}\ (\bibinfo  {publisher} {MIT press},\ \bibinfo {year}
  {2018})\BibitemShut {NoStop}%
\bibitem [{\citenamefont {Pope}(2001)}]{Pope_2001}%
  \BibitemOpen
  \bibfield  {author} {\bibinfo {author} {\bibfnamefont {S.~B.}\ \bibnamefont
  {Pope}},\ }\href {\doibase 10.1088/0957-0233/12/11/705} {\emph {\bibinfo
  {title} {Turbulent Flows}}},\ Vol.~\bibinfo {volume} {12}\ (\bibinfo
  {publisher} {{IOP} Publishing},\ \bibinfo {year} {2001})\ pp.\ \bibinfo
  {pages} {2020--2021}\BibitemShut {NoStop}%
\bibitem [{\citenamefont {Lamb}(1877)}]{Lamb_1877}%
  \BibitemOpen
  \bibfield  {author} {\bibinfo {author} {\bibfnamefont {H.}~\bibnamefont
  {Lamb}},\ }\bibfield  {title} {\enquote {\bibinfo {title} {{On the Conditions
  for Steady Motion of a Fluid}},}\ }\href {\doibase 10.1112/plms/s1-9.1.91}
  {\bibfield  {journal} {\bibinfo  {journal} {Proceedings of the London
  Mathematical Society}\ }\textbf {\bibinfo {volume} {s1-9}},\ \bibinfo {pages}
  {91--93} (\bibinfo {year} {1877})}\BibitemShut {NoStop}%
\bibitem [{\citenamefont {Banerjee}\ and\ \citenamefont
  {Galtier}(2016{\natexlab{a}})}]{Banerjee_2016c}%
  \BibitemOpen
  \bibfield  {author} {\bibinfo {author} {\bibfnamefont {S.}~\bibnamefont
  {Banerjee}}\ and\ \bibinfo {author} {\bibfnamefont {S.}~\bibnamefont
  {Galtier}},\ }\bibfield  {title} {\enquote {\bibinfo {title} {An alternative
  formulation for exact scaling relations in hydrodynamic and
  magnetohydrodynamic turbulence},}\ }\href {\doibase
  10.1088/1751-8113/50/1/015501} {\bibfield  {journal} {\bibinfo  {journal}
  {Journal of Physics A: Mathematical and Theoretical}\ }\textbf {\bibinfo
  {volume} {50}},\ \bibinfo {pages} {015501} (\bibinfo {year}
  {2016}{\natexlab{a}})}\BibitemShut {NoStop}%
\bibitem [{\citenamefont {Andr\'es}\ and\ \citenamefont
  {Banerjee}(2019)}]{Andres_2019b}%
  \BibitemOpen
  \bibfield  {author} {\bibinfo {author} {\bibfnamefont {N.}~\bibnamefont
  {Andr\'es}}\ and\ \bibinfo {author} {\bibfnamefont {S.}~\bibnamefont
  {Banerjee}},\ }\bibfield  {title} {\enquote {\bibinfo {title} {Statistics of
  incompressible hydrodynamic turbulence: An alternative approach},}\ }\href
  {\doibase 10.1103/PhysRevFluids.4.024603} {\bibfield  {journal} {\bibinfo
  {journal} {Phys. Rev. Fluids}\ }\textbf {\bibinfo {volume} {4}},\ \bibinfo
  {pages} {024603} (\bibinfo {year} {2019})}\BibitemShut {NoStop}%
\bibitem [{\citenamefont {Biskamp}(2003)}]{Biskamp_2003}%
  \BibitemOpen
  \bibfield  {author} {\bibinfo {author} {\bibfnamefont {D.}~\bibnamefont
  {Biskamp}},\ }\href {https://books.google.co.in/books?id=6iDZWv2DC-EC} {\emph
  {\bibinfo {title} {Magnetohydrodynamic Turbulence}}}\ (\bibinfo  {publisher}
  {Cambridge University Press},\ \bibinfo {year} {2003})\BibitemShut {NoStop}%
\bibitem [{\citenamefont {Galtier}(2016)}]{Galtier_2016}%
  \BibitemOpen
  \bibfield  {author} {\bibinfo {author} {\bibfnamefont {S.}~\bibnamefont
  {Galtier}},\ }\href {https://books.google.co.in/books?id=Idf4DAAAQBAJ} {\emph
  {\bibinfo {title} {Introduction to Modern Magnetohydrodynamics}}}\ (\bibinfo
  {publisher} {Cambridge University Press},\ \bibinfo {year}
  {2016})\BibitemShut {NoStop}%
\bibitem [{Note1()}]{Note1}%
  \BibitemOpen
  \bibinfo {note} {A range of intermediate length scales unaffected by both
  large-scale forcing and small-scale dissipation}\BibitemShut {NoStop}%
\bibitem [{\citenamefont {Banerjee}\ and\ \citenamefont
  {Galtier}(2013)}]{Banerjee_2013}%
  \BibitemOpen
  \bibfield  {author} {\bibinfo {author} {\bibfnamefont {S.}~\bibnamefont
  {Banerjee}}\ and\ \bibinfo {author} {\bibfnamefont {S.}~\bibnamefont
  {Galtier}},\ }\bibfield  {title} {\enquote {\bibinfo {title} {Exact relation
  with two-point correlation functions and phenomenological approach for
  compressible magnetohydrodynamic turbulence},}\ }\href {\doibase
  10.1103/PhysRevE.87.013019} {\bibfield  {journal} {\bibinfo  {journal} {Phys.
  Rev. E}\ }\textbf {\bibinfo {volume} {87}},\ \bibinfo {pages} {013019}
  (\bibinfo {year} {2013})}\BibitemShut {NoStop}%
\bibitem [{\citenamefont {Banerjee}\ and\ \citenamefont
  {Kritsuk}(2018)}]{Banerjee_2018}%
  \BibitemOpen
  \bibfield  {author} {\bibinfo {author} {\bibfnamefont {S.}~\bibnamefont
  {Banerjee}}\ and\ \bibinfo {author} {\bibfnamefont {A.~G.}\ \bibnamefont
  {Kritsuk}},\ }\bibfield  {title} {\enquote {\bibinfo {title} {Energy transfer
  in compressible magnetohydrodynamic turbulence for isothermal
  self-gravitating fluids},}\ }\href {\doibase 10.1103/PhysRevE.97.023107}
  {\bibfield  {journal} {\bibinfo  {journal} {Physical Review E}\ }\textbf
  {\bibinfo {volume} {97}},\ \bibinfo {pages} {023107} (\bibinfo {year}
  {2018})}\BibitemShut {NoStop}%
\bibitem [{\citenamefont {Banerjee}\ and\ \citenamefont
  {Andr{\'e}s}(2020)}]{Banerjee_2020}%
  \BibitemOpen
  \bibfield  {author} {\bibinfo {author} {\bibfnamefont {S.}~\bibnamefont
  {Banerjee}}\ and\ \bibinfo {author} {\bibfnamefont {N.}~\bibnamefont
  {Andr{\'e}s}},\ }\bibfield  {title} {\enquote {\bibinfo {title}
  {Scale-to-scale energy transfer rate in compressible two-fluid plasma
  turbulence},}\ }\href@noop {} {\bibfield  {journal} {\bibinfo  {journal}
  {Physical Review E}\ }\textbf {\bibinfo {volume} {101}},\ \bibinfo {pages}
  {043212} (\bibinfo {year} {2020})}\BibitemShut {NoStop}%
\bibitem [{\citenamefont {Verma}\ \emph {et~al.}(2021)\citenamefont {Verma},
  \citenamefont {Sharma}, \citenamefont {Chatterjee},\ and\ \citenamefont
  {Alam}}]{Verma_2021}%
  \BibitemOpen
  \bibfield  {author} {\bibinfo {author} {\bibfnamefont {M.}~\bibnamefont
  {Verma}}, \bibinfo {author} {\bibfnamefont {M.}~\bibnamefont {Sharma}},
  \bibinfo {author} {\bibfnamefont {S.}~\bibnamefont {Chatterjee}}, \ and\
  \bibinfo {author} {\bibfnamefont {S.}~\bibnamefont {Alam}},\ }\bibfield
  {title} {\enquote {\bibinfo {title} {Variable energy fluxes and exact
  relations in magnetohydrodynamics turbulence},}\ }\href {\doibase
  10.3390/fluids6060225} {\bibfield  {journal} {\bibinfo  {journal} {Fluids}\
  }\textbf {\bibinfo {volume} {6}} (\bibinfo {year} {2021}),\
  10.3390/fluids6060225}\BibitemShut {NoStop}%
\bibitem [{\citenamefont {Waleffe}(1992)}]{Waleffe_1992}%
  \BibitemOpen
  \bibfield  {author} {\bibinfo {author} {\bibfnamefont {F.}~\bibnamefont
  {Waleffe}},\ }\bibfield  {title} {\enquote {\bibinfo {title} {The nature of
  triad interactions in homogeneous turbulence},}\ }\href {\doibase
  10.1063/1.858309} {\bibfield  {journal} {\bibinfo  {journal} {Physics of
  Fluids A: Fluid Dynamics}\ }\textbf {\bibinfo {volume} {4}},\ \bibinfo
  {pages} {350--363} (\bibinfo {year} {1992})}\BibitemShut {NoStop}%
\bibitem [{\citenamefont {{Verma}}(2004)}]{Verma_2004}%
  \BibitemOpen
  \bibfield  {author} {\bibinfo {author} {\bibfnamefont {M.~K.}\ \bibnamefont
  {{Verma}}},\ }\bibfield  {title} {\enquote {\bibinfo {title} {{Statistical
  theory of magnetohydrodynamic turbulence: recent results}},}\ }\href
  {\doibase 10.1016/j.physrep.2004.07.007} {\bibfield  {journal} {\bibinfo
  {journal} {Phys. Rep.}\ }\textbf {\bibinfo {volume} {401}},\ \bibinfo {pages}
  {229--380} (\bibinfo {year} {2004})}\BibitemShut {NoStop}%
\bibitem [{\citenamefont {{Alexakis}}, \citenamefont {{Mininni}},\ and\
  \citenamefont {{Pouquet}}(2005)}]{Alexakis_2005}%
  \BibitemOpen
  \bibfield  {author} {\bibinfo {author} {\bibfnamefont {A.}~\bibnamefont
  {{Alexakis}}}, \bibinfo {author} {\bibfnamefont {P.~D.}\ \bibnamefont
  {{Mininni}}}, \ and\ \bibinfo {author} {\bibfnamefont {A.}~\bibnamefont
  {{Pouquet}}},\ }\bibfield  {title} {\enquote {\bibinfo {title}
  {{Shell-to-shell energy transfer in magnetohydrodynamics. I. Steady state
  turbulence}},}\ }\href {\doibase 10.1103/PhysRevE.72.046301} {\bibfield
  {journal} {\bibinfo  {journal} {Phys. Rev. E}\ }\textbf {\bibinfo {volume}
  {72}},\ \bibinfo {eid} {046301} (\bibinfo {year} {2005})}\BibitemShut
  {NoStop}%
\bibitem [{\citenamefont {{Mininni}}, \citenamefont {{Alexakis}},\ and\
  \citenamefont {{Pouquet}}(2005)}]{Mininni_2005b}%
  \BibitemOpen
  \bibfield  {author} {\bibinfo {author} {\bibfnamefont {P.}~\bibnamefont
  {{Mininni}}}, \bibinfo {author} {\bibfnamefont {A.}~\bibnamefont
  {{Alexakis}}}, \ and\ \bibinfo {author} {\bibfnamefont {A.}~\bibnamefont
  {{Pouquet}}},\ }\bibfield  {title} {\enquote {\bibinfo {title}
  {{Shell-to-shell energy transfer in magnetohydrodynamics. II. Kinematic
  dynamo}},}\ }\href {\doibase 10.1103/PhysRevE.72.046302} {\bibfield
  {journal} {\bibinfo  {journal} {Phys. Rev. E}\ }\textbf {\bibinfo {volume}
  {72}},\ \bibinfo {eid} {046302} (\bibinfo {year} {2005})}\BibitemShut
  {NoStop}%
\bibitem [{\citenamefont {Domaradzki}\ and\ \citenamefont
  {Rogallo}(1990)}]{Domaradzki_1990}%
  \BibitemOpen
  \bibfield  {author} {\bibinfo {author} {\bibfnamefont {J.~A.}\ \bibnamefont
  {Domaradzki}}\ and\ \bibinfo {author} {\bibfnamefont {R.~S.}\ \bibnamefont
  {Rogallo}},\ }\bibfield  {title} {\enquote {\bibinfo {title} {Local energy
  transfer and nonlocal interactions in homogeneous, isotropic turbulence},}\
  }\href {\doibase 10.1063/1.857736} {\bibfield  {journal} {\bibinfo  {journal}
  {Phys. Fluids A-Fluid}\ }\textbf {\bibinfo {volume} {2}},\ \bibinfo {pages}
  {413--426} (\bibinfo {year} {1990})}\BibitemShut {NoStop}%
\bibitem [{\citenamefont {{Kraichnan}}(1958)}]{Kraichnan_1958}%
  \BibitemOpen
  \bibfield  {author} {\bibinfo {author} {\bibfnamefont {R.~H.}\ \bibnamefont
  {{Kraichnan}}},\ }\bibfield  {title} {\enquote {\bibinfo {title}
  {{Irreversible Statistical Mechanics of Incompressible Hydromagnetic
  Turbulence}},}\ }\href {\doibase 10.1103/PhysRev.111.1747.4} {\bibfield
  {journal} {\bibinfo  {journal} {Phys. Rev.}\ }\textbf {\bibinfo {volume}
  {111}},\ \bibinfo {pages} {1747--1747} (\bibinfo {year} {1958})}\BibitemShut
  {NoStop}%
\bibitem [{\citenamefont {Sagaut}\ and\ \citenamefont
  {Cambon}(2008)}]{Sagaut_2008}%
  \BibitemOpen
  \bibfield  {author} {\bibinfo {author} {\bibfnamefont {P.}~\bibnamefont
  {Sagaut}}\ and\ \bibinfo {author} {\bibfnamefont {C.}~\bibnamefont
  {Cambon}},\ }\href@noop {} {\emph {\bibinfo {title} {Homogeneous turbulence
  dynamics}}},\ Vol.~\bibinfo {volume} {10}\ (\bibinfo  {publisher}
  {Springer},\ \bibinfo {year} {2008})\BibitemShut {NoStop}%
\bibitem [{\citenamefont {Lesieur}(1987)}]{Lesieur_1987}%
  \BibitemOpen
  \bibfield  {author} {\bibinfo {author} {\bibfnamefont {M.}~\bibnamefont
  {Lesieur}},\ }\href@noop {} {\emph {\bibinfo {title} {Turbulence in fluids:
  stochastic and numerical modelling}}},\ Vol.\ \bibinfo {volume} {488}\
  (\bibinfo  {publisher} {Nijhoff Boston, MA},\ \bibinfo {year}
  {1987})\BibitemShut {NoStop}%
\bibitem [{\citenamefont {Kraichnan}(1959)}]{Kraichnan_1959}%
  \BibitemOpen
  \bibfield  {author} {\bibinfo {author} {\bibfnamefont {R.~H.}\ \bibnamefont
  {Kraichnan}},\ }\bibfield  {title} {\enquote {\bibinfo {title} {The structure
  of isotropic turbulence at very high reynolds numbers},}\ }\href {\doibase
  10.1017/S0022112059000362} {\bibfield  {journal} {\bibinfo  {journal} {J.
  Fluid Mech.}\ }\textbf {\bibinfo {volume} {5}},\ \bibinfo {pages} {497–543}
  (\bibinfo {year} {1959})}\BibitemShut {NoStop}%
\bibitem [{\citenamefont {Dar}, \citenamefont {Verma},\ and\ \citenamefont
  {Eswaran}(2001)}]{DAR_2001}%
  \BibitemOpen
  \bibfield  {author} {\bibinfo {author} {\bibfnamefont {G.}~\bibnamefont
  {Dar}}, \bibinfo {author} {\bibfnamefont {M.~K.}\ \bibnamefont {Verma}}, \
  and\ \bibinfo {author} {\bibfnamefont {V.}~\bibnamefont {Eswaran}},\
  }\bibfield  {title} {\enquote {\bibinfo {title} {Energy transfer in
  two-dimensional magnetohydrodynamic turbulence: formalism and numerical
  results},}\ }\href {\doibase http://dx.doi.org/10.1016/S0167-2789(01)00307-4}
  {\bibfield  {journal} {\bibinfo  {journal} {Physica D}\ }\textbf {\bibinfo
  {volume} {157}},\ \bibinfo {pages} {207 -- 225} (\bibinfo {year}
  {2001})}\BibitemShut {NoStop}%
\bibitem [{\citenamefont {Plunian}, \citenamefont {Stepanov},\ and\
  \citenamefont {Verma}(2019)}]{Plunian_2019}%
  \BibitemOpen
  \bibfield  {author} {\bibinfo {author} {\bibfnamefont {F.}~\bibnamefont
  {Plunian}}, \bibinfo {author} {\bibfnamefont {R.}~\bibnamefont {Stepanov}}, \
  and\ \bibinfo {author} {\bibfnamefont {M.~K.}\ \bibnamefont {Verma}},\
  }\bibfield  {title} {\enquote {\bibinfo {title} {On uniqueness of transfer
  rates in magnetohydrodynamic turbulence},}\ }\href {\doibase
  10.1017/S0022377819000710} {\bibfield  {journal} {\bibinfo  {journal}
  {Journal of Plasma Physics}\ }\textbf {\bibinfo {volume} {85}},\ \bibinfo
  {pages} {905850507} (\bibinfo {year} {2019})}\BibitemShut {NoStop}%
\bibitem [{\citenamefont {Kumar}, \citenamefont {Verma},\ and\ \citenamefont
  {Samtaney}(2013)}]{Kumar_2013}%
  \BibitemOpen
  \bibfield  {author} {\bibinfo {author} {\bibfnamefont {R.}~\bibnamefont
  {Kumar}}, \bibinfo {author} {\bibfnamefont {M.~K.}\ \bibnamefont {Verma}}, \
  and\ \bibinfo {author} {\bibfnamefont {R.}~\bibnamefont {Samtaney}},\
  }\bibfield  {title} {\enquote {\bibinfo {title} {Energy transfers and
  magnetic energy growth in small-scale dynamo},}\ }\href {\doibase
  10.1209/0295-5075/104/54001} {\bibfield  {journal} {\bibinfo  {journal}
  {Europhysics Letters}\ }\textbf {\bibinfo {volume} {104}},\ \bibinfo {pages}
  {54001} (\bibinfo {year} {2013})}\BibitemShut {NoStop}%
\bibitem [{\citenamefont {Kumar}\ and\ \citenamefont
  {Verma}(2017)}]{Kumar_2017}%
  \BibitemOpen
  \bibfield  {author} {\bibinfo {author} {\bibfnamefont {R.}~\bibnamefont
  {Kumar}}\ and\ \bibinfo {author} {\bibfnamefont {M.~K.}\ \bibnamefont
  {Verma}},\ }\bibfield  {title} {\enquote {\bibinfo {title} {Amplification of
  large-scale magnetic field in nonhelical magnetohydrodynamics},}\ }\href
  {\doibase 10.1063/1.4997779} {\bibfield  {journal} {\bibinfo  {journal}
  {Physics of Plasmas}\ }\textbf {\bibinfo {volume} {24}},\ \bibinfo {pages}
  {092301} (\bibinfo {year} {2017})}\BibitemShut {NoStop}%
\bibitem [{\citenamefont {Halder}\ \emph {et~al.}(2023)\citenamefont {Halder},
  \citenamefont {Banerjee}, \citenamefont {Chatterjee},\ and\ \citenamefont
  {Sharma}}]{Halder_2023}%
  \BibitemOpen
  \bibfield  {author} {\bibinfo {author} {\bibfnamefont {A.}~\bibnamefont
  {Halder}}, \bibinfo {author} {\bibfnamefont {S.}~\bibnamefont {Banerjee}},
  \bibinfo {author} {\bibfnamefont {A.~G.}\ \bibnamefont {Chatterjee}}, \ and\
  \bibinfo {author} {\bibfnamefont {M.~K.}\ \bibnamefont {Sharma}},\ }\bibfield
   {title} {\enquote {\bibinfo {title} {Contribution of the hall term in
  small-scale magnetohydrodynamic dynamos},}\ }\href {\doibase
  10.1103/PhysRevFluids.8.053701} {\bibfield  {journal} {\bibinfo  {journal}
  {Phys. Rev. Fluids}\ }\textbf {\bibinfo {volume} {8}},\ \bibinfo {pages}
  {053701} (\bibinfo {year} {2023})}\BibitemShut {NoStop}%
\bibitem [{\citenamefont {Bittencourt}(2013)}]{Bittencourt_2013}%
  \BibitemOpen
  \bibfield  {author} {\bibinfo {author} {\bibfnamefont {J.}~\bibnamefont
  {Bittencourt}},\ }\href {https://books.google.co.in/books?id=XcreBwAAQBAJ}
  {\emph {\bibinfo {title} {Fundamentals of Plasma Physics}}}\ (\bibinfo
  {publisher} {Springer New York},\ \bibinfo {year} {2013})\BibitemShut
  {NoStop}%
\bibitem [{\citenamefont {Banerjee}\ and\ \citenamefont
  {Galtier}(2016{\natexlab{b}})}]{Banerjee_2016a}%
  \BibitemOpen
  \bibfield  {author} {\bibinfo {author} {\bibfnamefont {S.}~\bibnamefont
  {Banerjee}}\ and\ \bibinfo {author} {\bibfnamefont {S.}~\bibnamefont
  {Galtier}},\ }\bibfield  {title} {\enquote {\bibinfo {title} {Chiral exact
  relations for helicities in hall magnetohydrodynamic turbulence},}\ }\href
  {\doibase 10.1103/PhysRevE.93.033120} {\bibfield  {journal} {\bibinfo
  {journal} {Phys. Rev. E}\ }\textbf {\bibinfo {volume} {93}},\ \bibinfo
  {pages} {033120} (\bibinfo {year} {2016}{\natexlab{b}})}\BibitemShut
  {NoStop}%
\bibitem [{\citenamefont {Banerjee}\ and\ \citenamefont
  {Galtier}(2016{\natexlab{c}})}]{Banerjee_2016b}%
  \BibitemOpen
  \bibfield  {author} {\bibinfo {author} {\bibfnamefont {S.}~\bibnamefont
  {Banerjee}}\ and\ \bibinfo {author} {\bibfnamefont {S.}~\bibnamefont
  {Galtier}},\ }\bibfield  {title} {\enquote {\bibinfo {title} {An alternative
  formulation for exact scaling relations in hydrodynamic and
  magnetohydrodynamic turbulence},}\ }\href {\doibase
  10.1088/1751-8113/50/1/015501} {\bibfield  {journal} {\bibinfo  {journal}
  {Journal of Physics A: Mathematical and Theoretical}\ }\textbf {\bibinfo
  {volume} {50}},\ \bibinfo {pages} {015501} (\bibinfo {year}
  {2016}{\natexlab{c}})}\BibitemShut {NoStop}%
\bibitem [{\citenamefont {Banerjee}, \citenamefont {Halder},\ and\
  \citenamefont {Pan}(2023)}]{Banerjee_2023}%
  \BibitemOpen
  \bibfield  {author} {\bibinfo {author} {\bibfnamefont {S.}~\bibnamefont
  {Banerjee}}, \bibinfo {author} {\bibfnamefont {A.}~\bibnamefont {Halder}}, \
  and\ \bibinfo {author} {\bibfnamefont {N.}~\bibnamefont {Pan}},\ }\bibfield
  {title} {\enquote {\bibinfo {title} {Universal turbulent relaxation of fluids
  and plasmas by the principle of vanishing nonlinear transfers},}\ }\href
  {\doibase 10.1103/PhysRevE.107.L043201} {\bibfield  {journal} {\bibinfo
  {journal} {Phys. Rev. E}\ }\textbf {\bibinfo {volume} {107}},\ \bibinfo
  {pages} {L043201} (\bibinfo {year} {2023})}\BibitemShut {NoStop}%
\bibitem [{\citenamefont {Mininni}, \citenamefont {Alexakis},\ and\
  \citenamefont {Pouquet}(2007)}]{Mininni_2007}%
  \BibitemOpen
  \bibfield  {author} {\bibinfo {author} {\bibfnamefont {P.~D.}\ \bibnamefont
  {Mininni}}, \bibinfo {author} {\bibfnamefont {A.}~\bibnamefont {Alexakis}}, \
  and\ \bibinfo {author} {\bibfnamefont {A.}~\bibnamefont {Pouquet}},\
  }\bibfield  {title} {\enquote {\bibinfo {title} {Energy transfer in hall-mhd
  turbulence: cascades, backscatter, and dynamo action},}\ }\href {\doibase
  10.1017/S0022377806004624} {\bibfield  {journal} {\bibinfo  {journal}
  {Journal of plasma physics}\ }\textbf {\bibinfo {volume} {73}},\ \bibinfo
  {pages} {377--401} (\bibinfo {year} {2007})}\BibitemShut {NoStop}%
\bibitem [{\citenamefont {G{\'o}mez}, \citenamefont {Mininni},\ and\
  \citenamefont {Dmitruk}(2010)}]{Gomez_2010}%
  \BibitemOpen
  \bibfield  {author} {\bibinfo {author} {\bibfnamefont {D.~O.}\ \bibnamefont
  {G{\'o}mez}}, \bibinfo {author} {\bibfnamefont {P.~D.}\ \bibnamefont
  {Mininni}}, \ and\ \bibinfo {author} {\bibfnamefont {P.}~\bibnamefont
  {Dmitruk}},\ }\bibfield  {title} {\enquote {\bibinfo {title}
  {Hall-magnetohydrodynamic small-scale dynamos},}\ }\href {\doibase
  10.1103/PhysRevE.82.036406} {\bibfield  {journal} {\bibinfo  {journal}
  {Physical Review E}\ }\textbf {\bibinfo {volume} {82}},\ \bibinfo {pages}
  {036406} (\bibinfo {year} {2010})}\BibitemShut {NoStop}%
\bibitem [{\citenamefont {Miura}, \citenamefont {Yang},\ and\ \citenamefont
  {Gotoh}(2019)}]{Miura_2019}%
  \BibitemOpen
  \bibfield  {author} {\bibinfo {author} {\bibfnamefont {H.}~\bibnamefont
  {Miura}}, \bibinfo {author} {\bibfnamefont {J.}~\bibnamefont {Yang}}, \ and\
  \bibinfo {author} {\bibfnamefont {T.}~\bibnamefont {Gotoh}},\ }\bibfield
  {title} {\enquote {\bibinfo {title} {Hall magnetohydrodynamic turbulence with
  a magnetic prandtl number larger than unity},}\ }\href@noop {} {\bibfield
  {journal} {\bibinfo  {journal} {Physical Review E}\ }\textbf {\bibinfo
  {volume} {100}},\ \bibinfo {pages} {063207} (\bibinfo {year}
  {2019})}\BibitemShut {NoStop}%
\bibitem [{\citenamefont {Meyrand}\ \emph {et~al.}(2018)\citenamefont
  {Meyrand}, \citenamefont {Kiyani}, \citenamefont {G\"urcan},\ and\
  \citenamefont {Galtier}}]{Meyrand_2018}%
  \BibitemOpen
  \bibfield  {author} {\bibinfo {author} {\bibfnamefont {R.}~\bibnamefont
  {Meyrand}}, \bibinfo {author} {\bibfnamefont {K.~H.}\ \bibnamefont {Kiyani}},
  \bibinfo {author} {\bibfnamefont {O.~D.}\ \bibnamefont {G\"urcan}}, \ and\
  \bibinfo {author} {\bibfnamefont {S.}~\bibnamefont {Galtier}},\ }\bibfield
  {title} {\enquote {\bibinfo {title} {Coexistence of weak and strong wave
  turbulence in incompressible hall magnetohydrodynamics},}\ }\href {\doibase
  10.1103/PhysRevX.8.031066} {\bibfield  {journal} {\bibinfo  {journal} {Phys.
  Rev. X}\ }\textbf {\bibinfo {volume} {8}},\ \bibinfo {pages} {031066}
  (\bibinfo {year} {2018})}\BibitemShut {NoStop}%
\bibitem [{\citenamefont {Mouraya}\ and\ \citenamefont
  {Banerjee}(2019)}]{Mouraya_2019}%
  \BibitemOpen
  \bibfield  {author} {\bibinfo {author} {\bibfnamefont {S.}~\bibnamefont
  {Mouraya}}\ and\ \bibinfo {author} {\bibfnamefont {S.}~\bibnamefont
  {Banerjee}},\ }\bibfield  {title} {\enquote {\bibinfo {title} {Determination
  of energy flux rate in homogeneous ferrohydrodynamic turbulence using
  two-point statistics},}\ }\href {\doibase 10.1103/PhysRevE.100.053105}
  {\bibfield  {journal} {\bibinfo  {journal} {Phys. Rev. E}\ }\textbf {\bibinfo
  {volume} {100}},\ \bibinfo {pages} {053105} (\bibinfo {year}
  {2019})}\BibitemShut {NoStop}%
\bibitem [{\citenamefont {Pan}\ and\ \citenamefont
  {Banerjee}(2022)}]{Pan_2022}%
  \BibitemOpen
  \bibfield  {author} {\bibinfo {author} {\bibfnamefont {N.}~\bibnamefont
  {Pan}}\ and\ \bibinfo {author} {\bibfnamefont {S.}~\bibnamefont {Banerjee}},\
  }\bibfield  {title} {\enquote {\bibinfo {title} {Exact relations for energy
  transfer in simple and active binary fluid turbulence},}\ }\href {\doibase
  10.1103/PhysRevE.106.025104} {\bibfield  {journal} {\bibinfo  {journal}
  {Phys. Rev. E}\ }\textbf {\bibinfo {volume} {106}},\ \bibinfo {pages}
  {025104} (\bibinfo {year} {2022})}\BibitemShut {NoStop}%
\end{thebibliography}

\end{document}